\documentclass[aps,showpacs,11pt]{revtex4}
\usepackage{dcolumn}
\usepackage{graphicx}
\usepackage{amsmath}
\usepackage{amsfonts}
\usepackage{amssymb}
\usepackage{psfrag}
\usepackage{wrapfig}
\usepackage{subfigure}
\usepackage{makeidx}
\usepackage{bm}
\usepackage{epsf}
\usepackage{hyperref}
\usepackage{color}
\usepackage{multirow,dcolumn}
\usepackage{graphicx}

\begin{document}

\title{Complexity and phase transitions in a holographic QCD model}

\author{Shao-Jun Zhang}
\email{sjzhang84@hotmail.com}
\affiliation{Institute for Advanced Physics and Mathematics, Zhejiang University of Technology, Hangzhou 310023, China}
\date{\today}

\begin{abstract}

Applying the "Complexity=Action" conjecture, we study the holographic complexity close to crossover/phase transition in a holographic QCD model proposed by Gubser et al. This model can realize three types of phase transition, crossover or first and second order, depending on the parameters of the dilaton potential. The re-scaled late-time growth rate of holographic complexity density for the three cases is calculated. Our results show that it experiences a fast drop/jump close to the critical point while approaching constants far beyond the critical temperature. Moreover, close to the critical temperature, it shows a behavior characterizing the type of the transition. These features suggest that the growth rate of the holographic complexity may be used as a good parameter to characterize the phase transition. The Lloyd's bound is always satisfied for the cases we considered but only saturated for the conformal case.
\end{abstract}


\maketitle

\section{Introduction}

In recent years, AdS/CFT~\cite{Maldacena:1997re,Gubser:1998bc,Witten:1998qj} or the general holography duality has attracted lots of attention and efforts in community of physics. It has been successfully applied in various areas of modern physics, including QCD~\cite{Mateos:2007ay,Gubser:2009md,CasalderreySolana:2011us}, condensed matter physics (CMT)~\cite{Hartnoll:2009sz,Herzog:2009xv,McGreevy:2009xe,Horowitz:2010gk,Cai:2015cya}, cosmology~\cite{Banks:2004eb} and quantum information theory~\cite{Swingle:2009bg,Swingle:2012wq,Qi:2013caa}. Within this duality, quantities in the boundary field theories correspond to certain geometric quantities in the dual bulk gravitational theories. Recently, it is proposed that growth of quantum complexity of the boundary state is dual to the growth of the geometry of the interior of the black hole in the bulk~\cite{Susskind:2014rva}. In quantum field theory (or quantum information theory), complexity of a state is defined as the number of simple unitary operators (gates) needed to prepare the state from a reference state (vacuum for example). So this quantity measures the difficulty of producing the state of interest. Later, this proposal is refined into two concrete conjectures. One is known as "complexity=volume" (CV) conjecture~\cite{Stanford:2014jda}, which states that the complexity of the boundary state is equal to the volume of certain maximal spacelike slice ending at the boundary. The other most recent one is known as "complexity=action" (CA) conjecture~\cite{Brown:2015bva,Brown:2015lvg}, which states that the complexity of a boundary state is given by the on-shell bulk action on the so-called Wheeler-deWitt (WDW) patch, namely
\begin{eqnarray}\label{CA}
\mathcal{C} (\Sigma) = \frac{I_{\rm WDW}}{\pi \hbar},
\end{eqnarray}
where $\Sigma$ is the time slice of the boundary where the QFT lives, and the WDW patch is the domain of dependence of some Cauchy surface in the bulk ending on $\Sigma$ at the boundary. It is also conjectured that the growth rate of the complexity satisfies the Lloyd's bound~\cite{Lloyd:2000}, that is
\begin{eqnarray}\label{LloydBound}
\frac{d \mathcal{C}}{d t} \leq \frac{2 M}{\pi},
\end{eqnarray}
where $M$ is the average energy of the state at time $t$, which is dual to the energy of the bulk state (for example the black hole mass). And it is saturated for neutral cases. Inspired by these works, there raises an intensive interest in studying the holographic complexity and the Lloyd's bound for various holographic gravity models~\cite{Alishahiha:2015rta,Momeni:2016ekm,Cai:2016xho,Brown:2016wib,Couch:2016exn,Yang:2016awy,Chapman:2016hwi,
Carmi:2016wjl,Pan:2016ecg,Brown:2017jil,Kim:2017lrw,Cai:2017sjv,Alishahiha:2017hwg,Bakhshaei:2017qud,
Tao:2017fsy,Guo:2017rul,Zangeneh:2017tub,Alishahiha:2017cuk,Abad:2017cgl,Reynolds:2017lwq,Hashimoto:2017fga,Nagasaki:2017kqe,Miao:2017quj,Ge:2017rak,
Ghodrati:2017roz,Qaemmaqami:2017lzs,Carmi:2017jqz,Kim:2017qrq,Cottrell:2017ayj,Sebastiani:2017rxr,
Moosa:2017yvt,HosseiniMansoori:2017tsm,Reynolds:2017jfs}.

In this paper, we aim at studying the properties of complexity, based on the CA conjecture, in the holographic QCD model proposed by Gubser and his collaborators~\cite{Gubser:2008ny,Gubser:2008yx}. As we have known for a long time that QCD is a strongly coupled theory at low energy, so it is difficult and a challenge to study its low energy physic, such as confinement/deconfinement transition or etc, with traditional methods. As a strong/weak duality, AdS/CFT may provide us powerful tools to deal with strongly coupled problems in quantum field theory by mapping them to simpler classical problems in the dual gravitational theory. In Gubser's holographic QCD model, beyond the usual Einstein gravity sector, a non-trivial dilaton field as well as a judicious choice of its potential are introduced to break the conformal symmetry intending to mimic the equation of state of the real QCD. The specific dilaton potential is parameterized by several constants, and by choosing appropriate values of them, the model can exhibit a crossover behavior at some critical temperature and the thermodynamical properties (such as the speed of sound) generated agree well with the results from lattice QCD (lQCD)~\cite{Borsanyi:2012cr}. Moreover, by choosing other appropriate values of the parameters in the dilaton potential, this simple model can also realize first and second order phase transitions. Thus, in additional to mimic properties of the real QCD, this model can also provide us a good background to study various phase structures of strongly coupled field systems within the same framework. Many efforts have been devoted to investigate properties of this model. In Ref.~\cite{Finazzo:2014cna}, various hydrodynamic transport coefficients are calculated. In Refs.~\cite{Janik:2015waa,Janik:2016btb}, quasinormal modes (QNMs) are calculated to probe the crossover/phase transition and a number of novel features are observed which were absent in the conformal case. In Ref.~\cite{DeWolfe:2010he}, this model is also extended to include the effect of finite chemical potential by including the Maxwell field in the bulk, and thus a more complete phase diagram of QCD is studied.

In Ref.~\cite{Zhang:2016rcm}, we apply a non-local observable, the holographic entanglement entropy (HEE), to probe the phase structures in the Gubser's model for the three cases. It is found that HEE experiences a fast drop or jump when the temperature approaching the critical point which can be seen as a signal of confinement. Moreover, close to the critical point, HEE shows behaviors characterizing the type of the transition. These properties of HEE suggest that we may use it to characterize the corresponding phase structures. This work has been extended to including the effect of the chemical potential in Ref.~\cite{Knaute:2017lll}. As there is a deep connection between the HEE and holographic complexity, it is natural to study the critical behavior of complexity in this model to see if it can also characterize the corresponding phase structures. Indeed, as we will show in the main context, holographic complexity shows a novel behavior suggesting that it is a good parameter to describe the corresponding phase structures.

The paper is organized as follows. In the next section, we will give a brief introduction of the holographic QCD model proposed by Guber et al. Then, in Sec. III, thermodynamics of the model, especially the entropy density, is discussed. In Sec. IV, behaviors of the late-time growth of the holographic complexity close to the crossover/phase transitions is investigated. The last section is devoted to summary and discussions.

\section{Review of the holographic QCD model}

We will consider the holographic QCD model proposed by Gubser et al in Refs.~\cite{Gubser:2008ny,Gubser:2008yx}. The model is a Einstein-dilaton theory with the full bulk action
\begin{eqnarray}\label{BulkAction}
S = \frac{1}{16 \pi G_5}\int d^5 x \sqrt{-g} \left[R-\frac{1}{2} (\partial \phi)^2 - V(\phi)\right] + \frac{1}{8 \pi G_5} \int_{\partial \mathcal{M}} K,
\end{eqnarray}
where the last term is the well-known York-Gibbons-Hawking (YGH) surface term which is required to make the variation problem well-defined, and $K$ is the trace of the extrinsic curvature of the boundary. $G_5$ is the gravitational constant, and we will apply the conventions that $16 \pi G_5 =1$ and $\hbar = 1$ in the following discussions. To mimic the thermodynamics properties of the real QCD, the authors assumes that the dilaton potential takes a specific form~\cite{Janik:2015waa,Janik:2016btb}
\begin{eqnarray}\label{ScalarPotential}
V(\phi)= -12  \cosh (\gamma \phi) + b_2 \phi^2 + b_4 \phi^4 + b_6 \phi^6,
\end{eqnarray}
where $\gamma, b_2, b_4$ and $b_6$ are four parameters whose values we can choose. It has the following perturbation expansion at small $\phi$
\begin{eqnarray}
V(\phi) \sim  -12 + \frac{1}{2} m^2 \phi^2 + {\cal O} (\phi^4).
\end{eqnarray}
The first term is the standard negative cosmological constant (note that we have chosen the unit to set the AdS radius to be one), and the second term is the mass term with $m^2 \equiv 2(b_2-6 \gamma^2)$. According to the AdS/CFT dictionary, the dilaton field $\phi$ in the bulk is dual to a scalar operator $O_\phi$ in the dual boundary field theory. The conformal dimension $\Delta$ of $O_\phi$ is determined by the mass as $\Delta = 2 \pm \sqrt{m^2 +4}$. The mass square $m^2$ can be negative and is constrained by the Breitenloner-Freedman (BF) bound $m^2 \geq -4$~\cite{Breitenlohner:1982bm,Breitenlohner:1982jf}. Actually, including the non-trivial dilaton field in the bulk is dual to making a deformation of the boundary conformal field theory (CFT)
\begin{eqnarray}
{\cal L} = {\cal L}_{\rm CFT} + E^{4-\Delta} O_{\phi},
\end{eqnarray}
where $E$ is an energy scale. In this paper, we only consider $2 \leq \Delta <4$ which corresponds to relevant deformations of the CFT.

By choosing suitable values of the parameters $(\gamma, b_2, b_4, b_6)$ in the dilaton potential, this model can produce thermodynamical properties, for example the temperature-dependence of the entropy density and the speed of sound, which agree well with the lQCD data. Moreover, by choosing other appropriate values of the parameters, this model can also realize various types of phase transitions. In this work, as in Refs.~\cite{Janik:2015waa,Janik:2016btb,Zhang:2016rcm} and shown in Table~1, we will consider three sets of parameters, labeled by $V_{\rm QCD}, V_{\rm 1st}$ and $V_{\rm 2nd}$ respectively. The parameters for $V_{\rm QCD}$ are chosen to fit the lQCD data from Ref.~\cite{Borsanyi:2012cr}, and the system is known to possess a crossover behaviour at certain critical temperature. Parameters of potentials $V_{\rm 1st}$ and $V_{\rm 2nd}$ are chosen so that the corresponding dual field systems exhibit respectively the $1^{\rm st}$, and the $2^{\rm nd}$ order phase transitions at certain critical temperature.

\begin{table}[!htbp]
\begin{tabular}{c c c c c c c}
\hline
\hline
potential &  $\gamma$ & $b_2$ & $b_4$ &  $b_6$ &  $\Delta$ \\
\hline
$V_{\rm QCD}$ &  0.606 & 1.4 & -0.1 & 0.0034 & 3.55\\
$V_{\rm 2nd}$ &  $1/\sqrt{2}$ & 1.958 & 0 & 0 & 3.38 \\
$V_{\rm 1st}$ &  $\sqrt{7/12}$ & 2.5 & 0 & 0 & 3.41\\
\hline
\end{tabular}
\caption{Parameters for the three scalar potentials~\cite{Janik:2016btb}.}
\end{table}

From the action, we can derive the equations of motion
\begin{eqnarray}\label{EOM}
R_{\mu\nu} - \frac{1}{2} g_{\mu\nu} R &=& \frac{1}{2} \left[\partial_\mu \phi \partial_\nu \phi -\frac{1}{2} g_{\mu\nu} (\partial \phi)^2 - g_{\mu\nu} V(\phi)\right],\nonumber\\
\nabla^2 \phi &=& \frac{\partial V}{\partial \phi}.
\end{eqnarray}
As we want to study properties of the dual field system at finite temperature, we need black hole solutions in the gravity side. To seek these solutions, we consider the following ansatz which takes the Gubser's gauge as in Refs.~\cite{Gubser:2008ny,Gubser:2008yx},
\begin{eqnarray}\label{ansatz}
ds^2 &=& e^{2 A} (-h dt^2 + d\vec{x}^2) + \frac{e^{2 B}}{h} dz^2.\nonumber\\
\phi &=& z,
\end{eqnarray}
where $A, B$ and $h$ are only functions of the radial coordinate $z$ (or, equivalently $\phi$). The infinite asymptotical boundary is $z \rightarrow 0$ and the singularity at $z\rightarrow \infty$, and the horizon $\phi=z_H$ is determined by the zero point of the blackening function $h$:
\begin{eqnarray}\label{horizon}
h(z_H)=0.
\end{eqnarray}
It is easy to see that the metric (\ref{ansatz}) recovers the standard Schwarzschild-AdS black hole when
\begin{eqnarray}\label{SAdS}
A = B = -\ln z, \quad h(z) &=& 1 - \left(\frac{z}{z_H}\right)^4, \quad \phi=0,
\end{eqnarray}
which corresponds to the conformal case. Generally, it is difficult to get analytical solutions with non-trivial dilaton configurations, so we follow the numerical strategy proposed in Refs.~\cite{Gubser:2008ny,Gubser:2008yx} to solve the field equations. Briefly, given one value of the horizon $z_H$, one unique black hole solution can be obtained. In this paper, we will vary the value of $z_H$ and obtain a family of black hole solutions numerically, so that we can study the temperature-dependence of various physical observable.

\section{Thermodynamics}

In this section, with the numerical black hole solutions, we study the thermodynamics of the dual field system holographically. We will focus on the temperature-dependence of the entropy density.

From the ansatz Eq.~(\ref{ansatz}) and in the units $16 \pi G_5 =1$ and $\hbar=1$, the Hawking temperature and the entropy density can be obtained,
\begin{eqnarray}
T = \frac{e^{A(z_H) -B(z_H)} |h'(z_H)|}{4\pi}, \qquad s = 4 \pi e^{3 A(z_H)}.
\end{eqnarray}
For the conformal case (\ref{SAdS}), restoring the unit and using the AdS/CFT dictionary, it is easy to derive the following relation
\begin{eqnarray}
\frac{s}{T^3} = \frac{4 \pi^4}{16 \pi G_5} = \frac{\pi^2}{2} N_c^2,
\end{eqnarray}
where $N_c$ is the group rank of the dual CFT and thus $N_c^2$ can be seen as a rough count of the number of degrees of freedom.\\
Using the numerical solutions, we show the temperature dependence of $s/T^3$ for the three non-conformal cases in Fig.~1.

\begin{figure}[!htbp]
\centering
\subfigure[$~~V_{\rm QCD}$]{\includegraphics[width=0.3\textwidth]{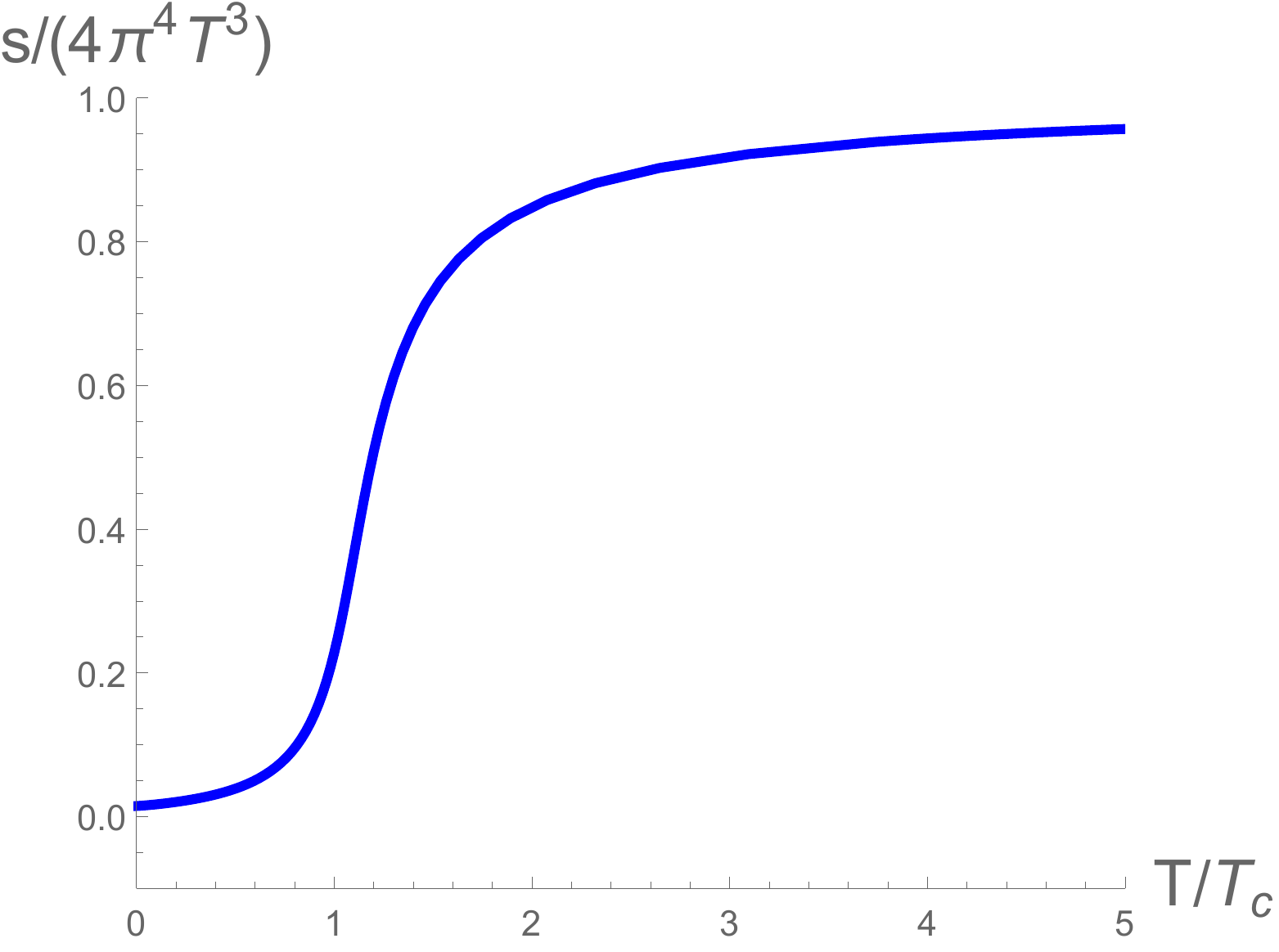}}\quad
\subfigure[$~~V_{\rm 1st}$]{\includegraphics[width=0.3\textwidth]{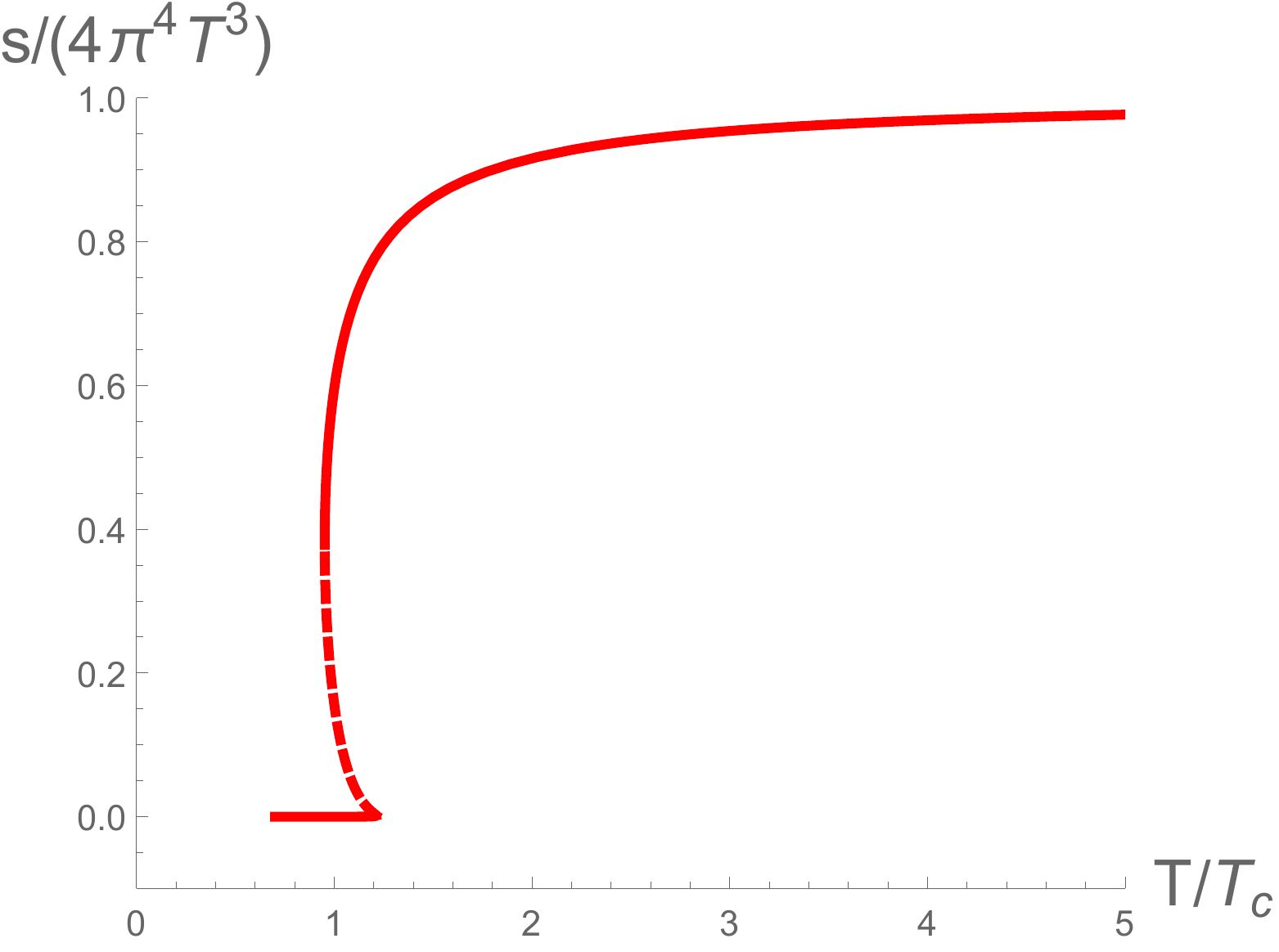}}\quad
\subfigure[$~~V_{\rm 2nd}$]{\includegraphics[width=0.3\textwidth]{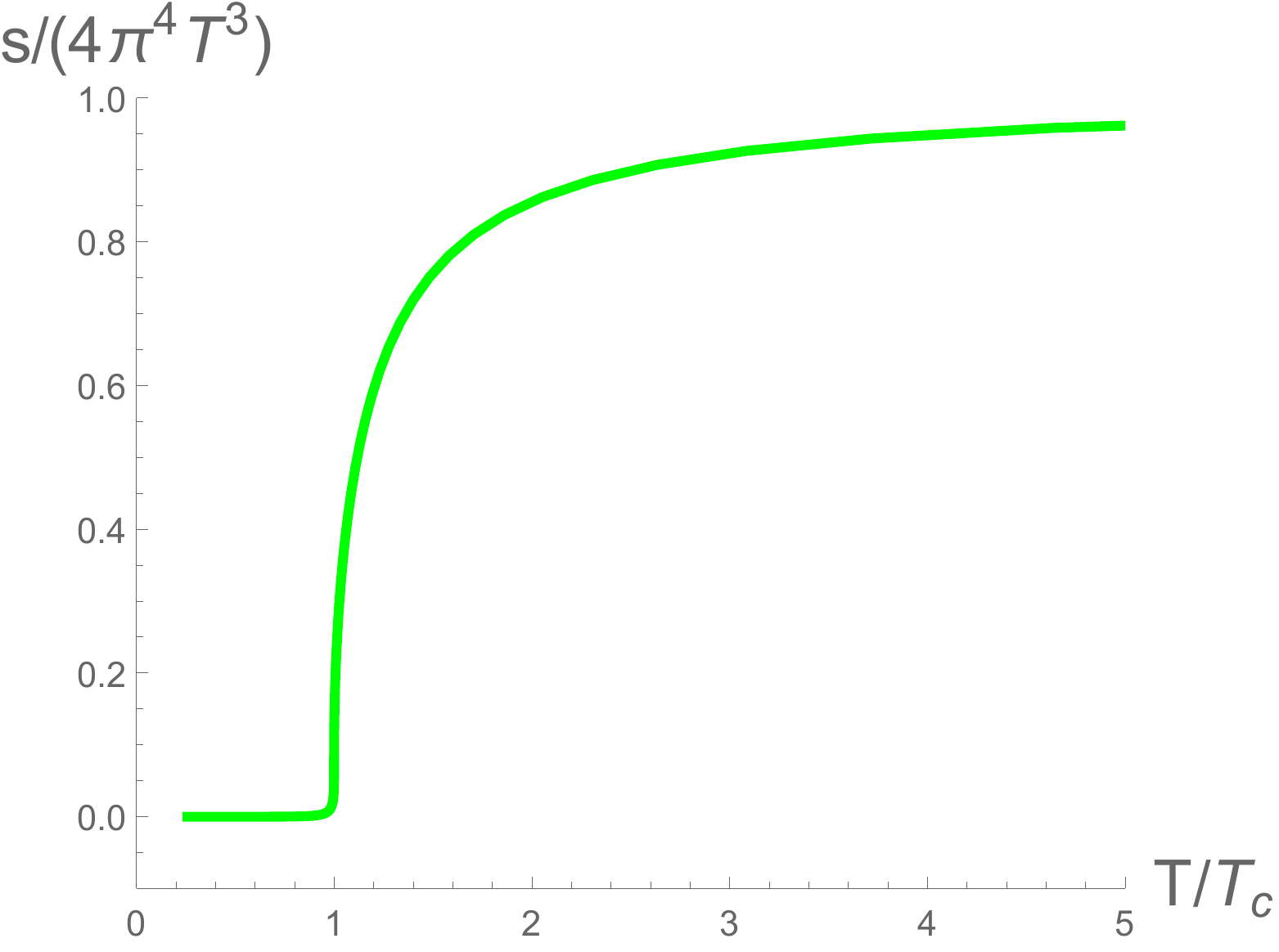}}
\caption{(color online) Temperature-dependence of entropy density for the three non-conformal cases. The red dashed line in the middle panel denotes the unstable solutions.}
\end{figure}

For the case $V_{\rm QCD}$, thermodynamical properties, for example the square of the speed of sound $c_s^2$, produced fit well with the lattice results~\cite{Borsanyi:2012cr,Janik:2015waa,Zhang:2016rcm}. The critical temperature $T_c$ is defined by the lowest dip of $c_s^2$, which in our present unit is $T_c=0.181033$. See more details, please refer to Ref.~\cite{Zhang:2016rcm}. As shown in the left panel, the re-scaled entropy density $s/T^3$ experiences a crossover at the critical temperature while approaching constant far beyond the critical point. This property suggests that even in the non-conformal case, $s/T^3$ may be considered to approximately count the effective number of degrees of freedom (DOF), namely we may have the relation
\begin{eqnarray}\label{soverT3}
\frac{s}{T^3} \propto {\rm effective\ number\ of\ DOF}.
\end{eqnarray}
As we will see later, this claim is also true for the other two cases, $V_{\rm 1st}$ and $V_{\rm 2nd}$. Moreover, as the temperature approaching $T_c$, the re-scaled entropy density $s/T^3$ drops quickly indicating that the number of underlying degrees of freedom is largely suppressed, which can be understood as a signal of confinement.

In the middle panel for the case $V_{\rm 1st}$, we can see that in some temperature range, three branches of solutions co-exist, in which two of them are stable (shown in solid curves) and one is unstable (shown in dashed curve).
By comparing the free energy of the two stable branches of solutions, it is found that there is a first-order phase transition in between at critical temperature $T_c =0.243901$~\cite{Janik:2016btb}. Moreover, we can see that $s/T^3$ jumps as the temperature approaching $T_c$ which also indicates that the number of underlying degrees of freedom is largely suppressed suddenly, and can also be seen as a kind of confinement. However, we should note that for the two cases $V_{\rm 1st}$ and $V_{\rm 2nd}$, we do not intend to mimic the real QCD, but rather realize various types of phase structures within the same framework.

In the right panel for $V_{\rm 2nd}$, there is a second-order phase transition with the critical temperature corresponding to the zero point of the speed of sound $c_s$. In our present units, $T_c=0.156841$. And once again, $s/T^3$ drops quickly as the temperature approaching $T_c$.

\section{Growth rate of holographic complexity}

In this section, we will study the behaviors of complexity holographically for the three non-conformal cases close to the transition.

According to CA conjecture Eq.~(\ref{CA}), to calculate the complexity of a boundary state, we need to calculate the on-shell bulk action on the WDW patch. Rigorous calculation of the on-shell action on this patch requires careful treatment of the contributions from the null boundaries of the patch and their joints~\cite{Lehner:2016vdi}.  In Ref.~\cite{Brown:2015lvg}, the authors argue that the growth rate of the holographic complexity at late time can be given by evaluating the on-shell action in the spacetime region inside the horizon $z=z_H$ (including the contributions from the two boundaries $z=z_H$ and $z=\infty$). In this work, as we only concentrate on the growth rate of the holographical complexity at late time, we can follow this strategy to do the calculations without knowing too much details of the contributions from the null boundaries of the WDW patch and their joints.

By using the equations of motion Eq.~(\ref{EOM}), in the late-time interval $[t, t+\delta t]$, the bulk term of the on-shell action inside the horizon is
\begin{eqnarray}
S_{\rm bulk} &=& \int d^5 x \ \frac{2}{3} e^{4 A + B} V(z) \nonumber\\
&=& \frac{2 V_3}{3} \int_{t}^{t+\delta t} dt \int_{z_H}^{\infty} e^{4 A + B} V(z) dz,
\end{eqnarray}
and the YGH term is
\begin{eqnarray}
S_{\rm YGH} = \delta t V_3 \bigg[e^{-4 A - B} \partial_r \big(e^{8 A} h\big)\bigg]\Bigg|_{z_H}^\infty,
\end{eqnarray}
where $V_3 \equiv \int d^3\vec{x}$ is the volume of the boundary field system. So, the growth rate of the holographic complexity density $c \equiv \mathcal{C}/V_3$ at late time is
\begin{eqnarray}
\frac{d c}{d t} =  \frac{2}{3 \pi} \int_{z_H}^{\infty} e^{4 A + B} V(z) dz +\frac{1}{\pi} \bigg[e^{-4 A - B} \partial_r \big(e^{8 A} h\big)\bigg]\Bigg|_{z_H}^\infty.\label{RateHoloCompDensity}
\end{eqnarray}
As we rely on numerical calculations, we can not set $z$ to infinity in practice. Actually, we set $z$ to a cutoff $z_c$ which is very big. It is found that the behavior of $dc/dt$ does not affected by the exact value of the cutoff which demonstrates the reliability of our calculations.

\subsection{Conformal case: Schwarzschild-AdS black hole}

To make a comparison with the conformal case, here we also show the results for the conformal case Eq.~(\ref{SAdS}). Using Eq.~(\ref{RateHoloCompDensity}), the growth rate of the holographic complexity density at late time is
\begin{eqnarray}
\frac{d c}{d t} = \frac{6}{\pi z_H^4}.
\end{eqnarray}
It is easy to see that $\frac{d c}{dt} = \frac{3}{2\pi} T s$. Recalling that the energy density of the black hole is $\varepsilon = \frac{3}{z_H^4}$, we have $d c/d t = 2 \varepsilon/\pi$ saturating the Lloyd's bound Eq.~(\ref{LloydBound}). Considering that the Hawking temperature is $T = \frac{1}{\pi z_H}$, it can also be written as
\begin{eqnarray}
\frac{d c/dt}{6 \pi^3 T^4} = 1.
\end{eqnarray}
Restoring the unit, it can be written as $\frac{d c/dt}{6 \pi^3 T^4} = \frac{1}{16 \pi G_5} \propto N_c^2$. So the re-scaled growth rate of the complexity density $\frac{d c/dt}{T^4}$ may also be seen as an approximate count of the degrees of freedom of the boundary CFT.

\subsection{The three non-conformal cases}

In Fig.~2, we show the temperature dependence of the re-scaled growth rate of the holographic complexity density for the three non-conformal cases.

\begin{figure}[!htbp]
\centering
\subfigure[$~~V_{\rm QCD}$]{\includegraphics[width=0.3\textwidth]{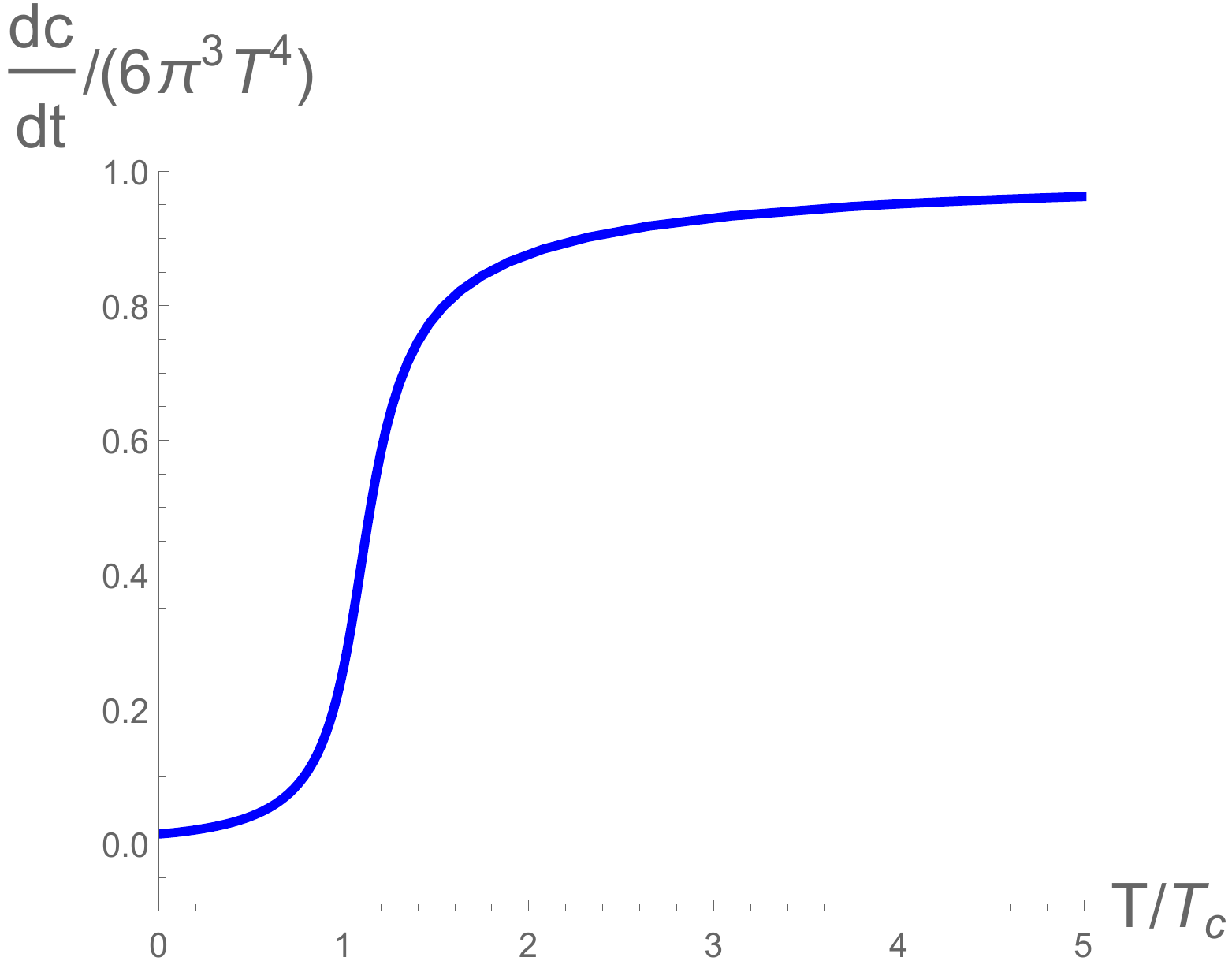}}\quad
\subfigure[$~~V_{\rm 1st}$]{\includegraphics[width=0.3\textwidth]{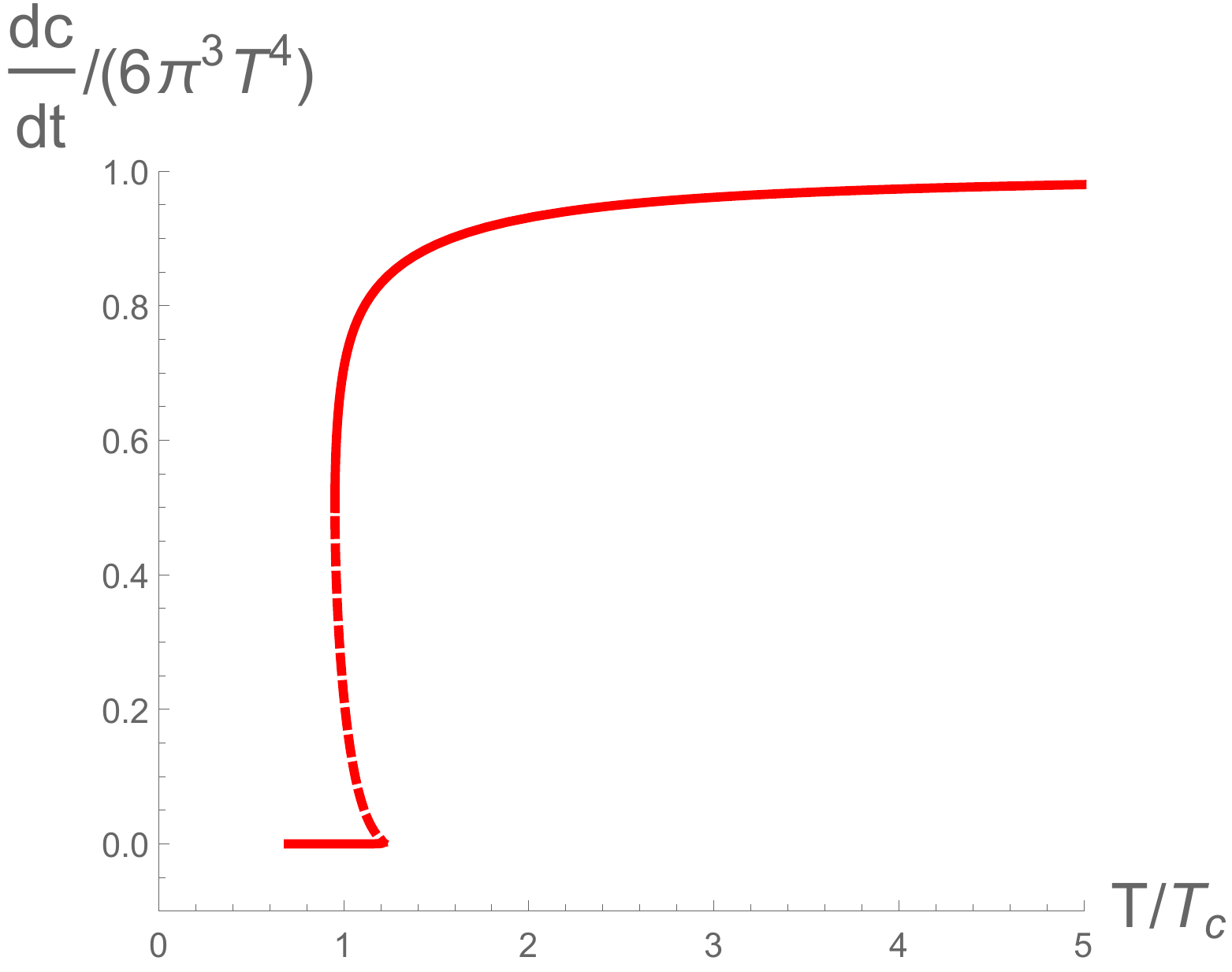}}\quad
\subfigure[$~~V_{\rm 2nd}$]{\includegraphics[width=0.3\textwidth]{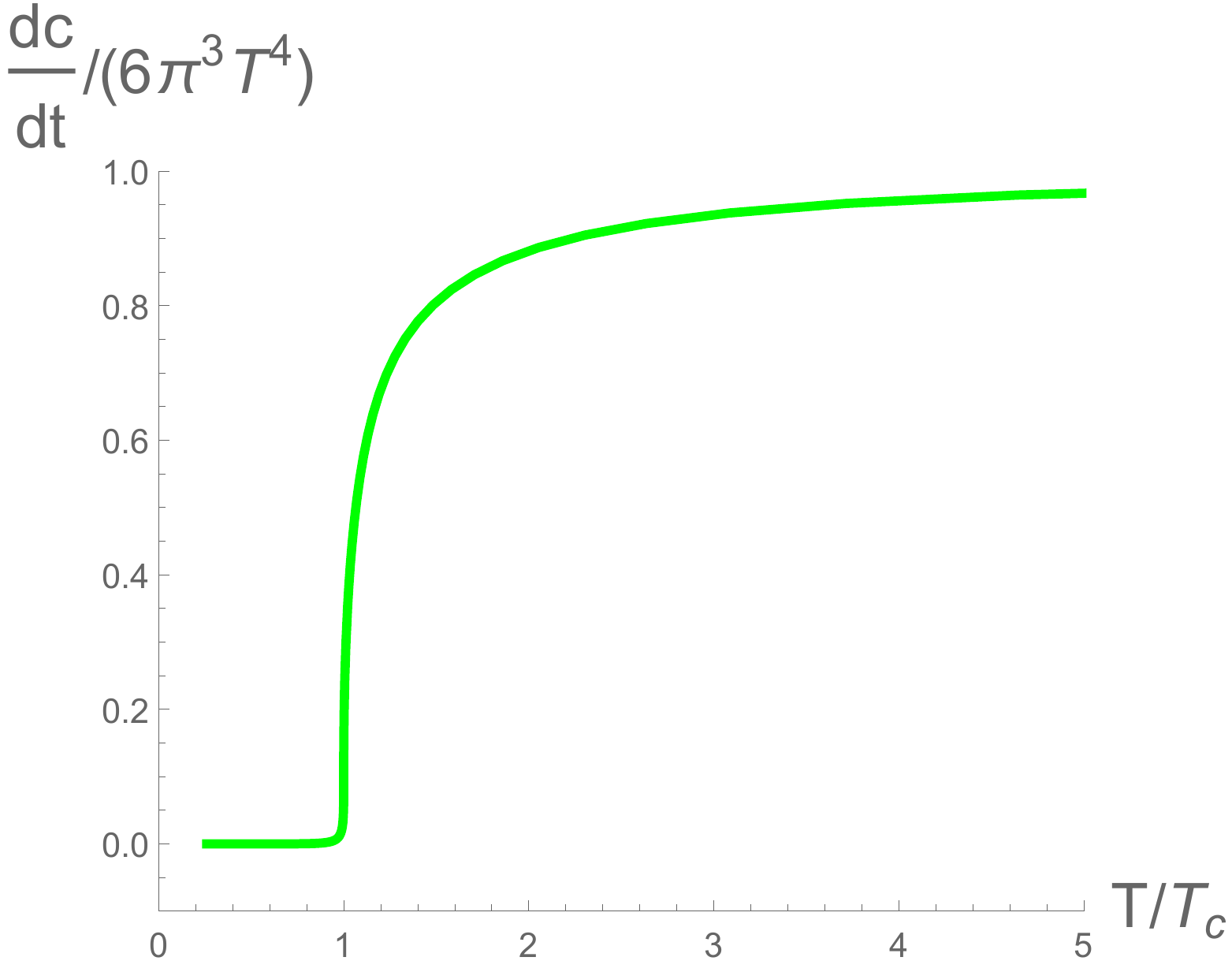}}
\caption{(color online) Temperature dependence of the growth rate of holographic complexity density for the three non-conformal cases.}
\end{figure}

From the figure, we can see that for the three non-conformal cases, the re-scaled growth rate of the holographic complexity density exhibits behavior characterizing the corresponding phase structures close to $T_c$: For the crossover case $V_{\rm QCD}$, $\frac{dc/dt}{T^4}$ and its derivative with respect to the temperature are both continuous at $T_c$; For the $1^{\rm st}$ order case $V_{\rm 1st}$, $\frac{dc/dt}{T^4}$ is discontinuous at $T_c$; For the $2^{\rm nd}$ order case $V_{\rm 2nd}$, $\frac{dc/dt}{T^4}$ is continuous at $T_c$ but not its derivative with respect to the temperature; While far beyond the critical point, it approaches zero as the temperature approaching zero and approaches the conformal value as the temperature becomes high enough. This is a little different from the behavior of holographic entanglement entropy~\cite{Zhang:2016rcm} which does not approach constant far beyond the critical temperature.

Similar to the conformal case, it can be checked that we still have the relation $\frac{d c}{dt} \sim T s$ so $\frac{d c/dt}{T^4} \sim  B(T) N_c^2$, but now the proportional factor $B(T)$ is temperature-dependent. From Fig.~2, we can see that $B(T)$ approaches its conformal constant value when the temperature is high enough while approaches zero when the temperature is low enough. So $B(T) N_c^2$ (or $\frac{d c/dt}{T^4}$) may be seen as the temperature-dependent effective number of degrees of freedom (DOF). Namely, similar to Eq.~(\ref{soverT3}), we may also have the relation
\begin{eqnarray}\label{dcdt}
\frac{dc/dt}{T^4} \propto {\rm effective\ number\ of\ DOF}.
\end{eqnarray}
And as the re-scaled thermal entropy density $s/T^3$, $\frac{dc/dt}{T^4}$ may also be used as a good parameter to characterize the type of phase transition. So, the fast drop or jump of $\frac{dc/dt}{T^4}$ near the critical temperature can be physically understood as follows: as the temperature approaching $T_c$, the field system undergoes confinement so that the number of degrees of freedom contributing to the complexity is largely suppressed.

\begin{figure}[!htbp]
\centering
\subfigure[$~~V_{\rm QCD}$]{\includegraphics[width=0.3\textwidth]{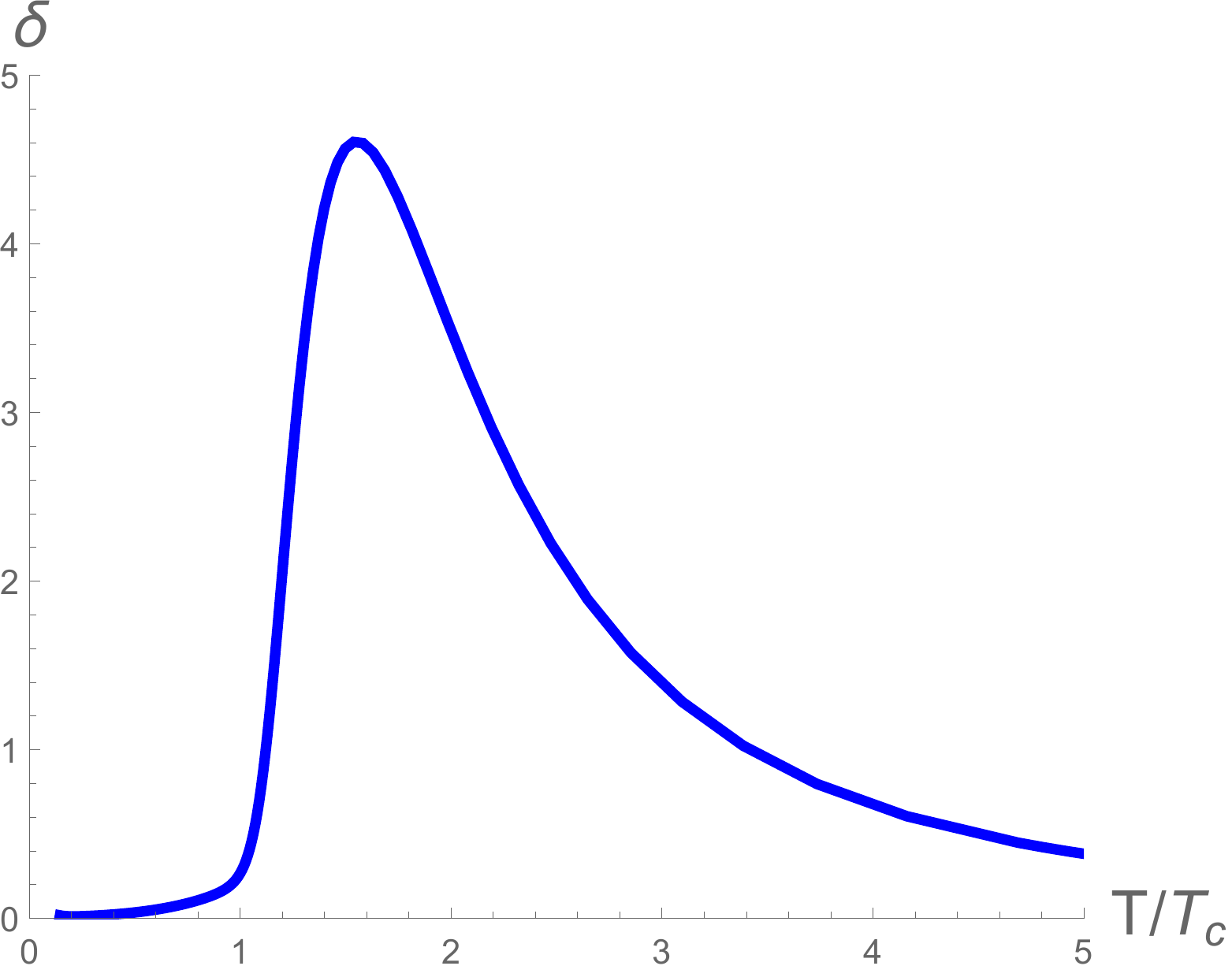}}\quad
\subfigure[$~~V_{\rm 1st}$]{\includegraphics[width=0.3\textwidth]{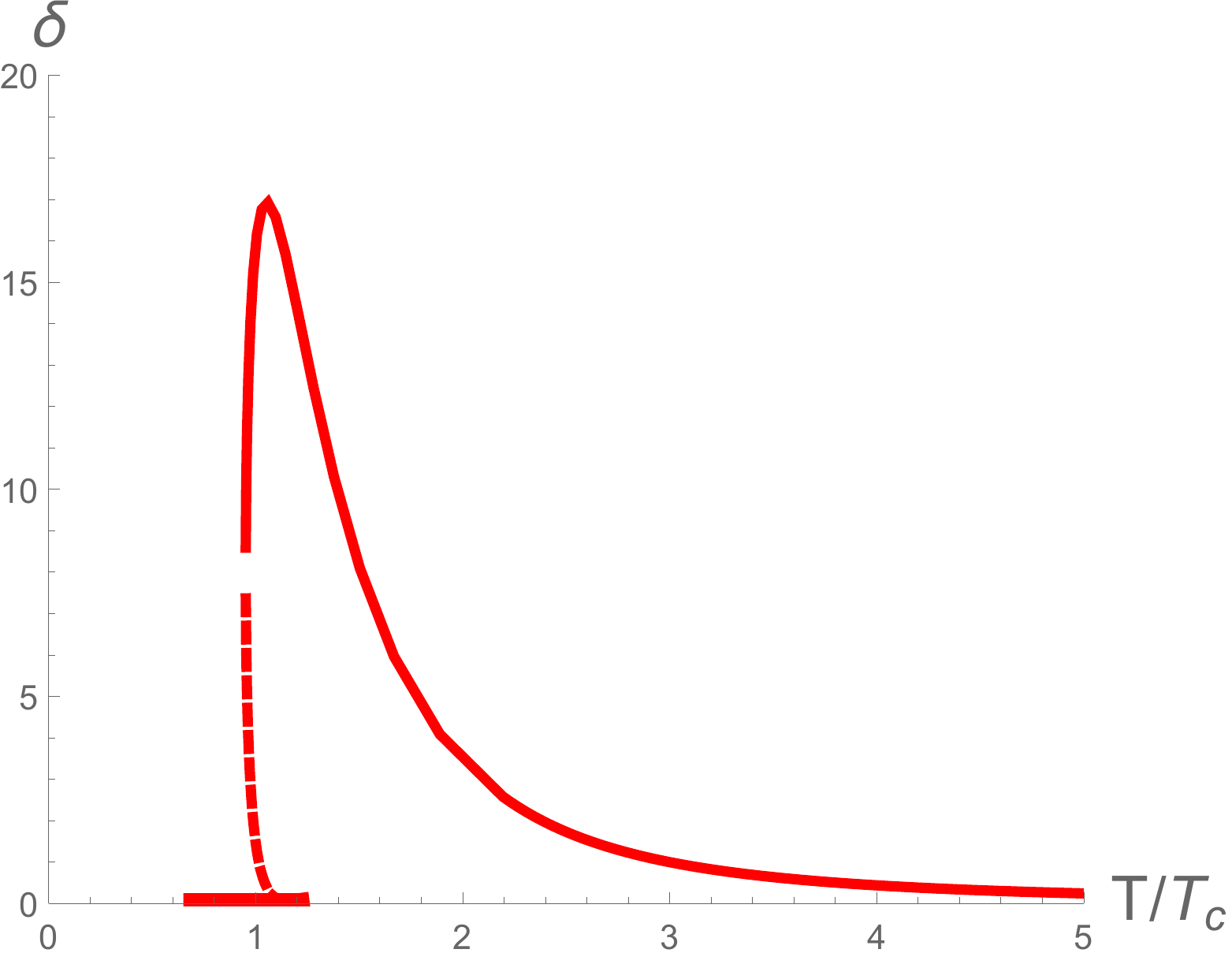}}\quad
\subfigure[$~~V_{\rm 2nd}$]{\includegraphics[width=0.3\textwidth]{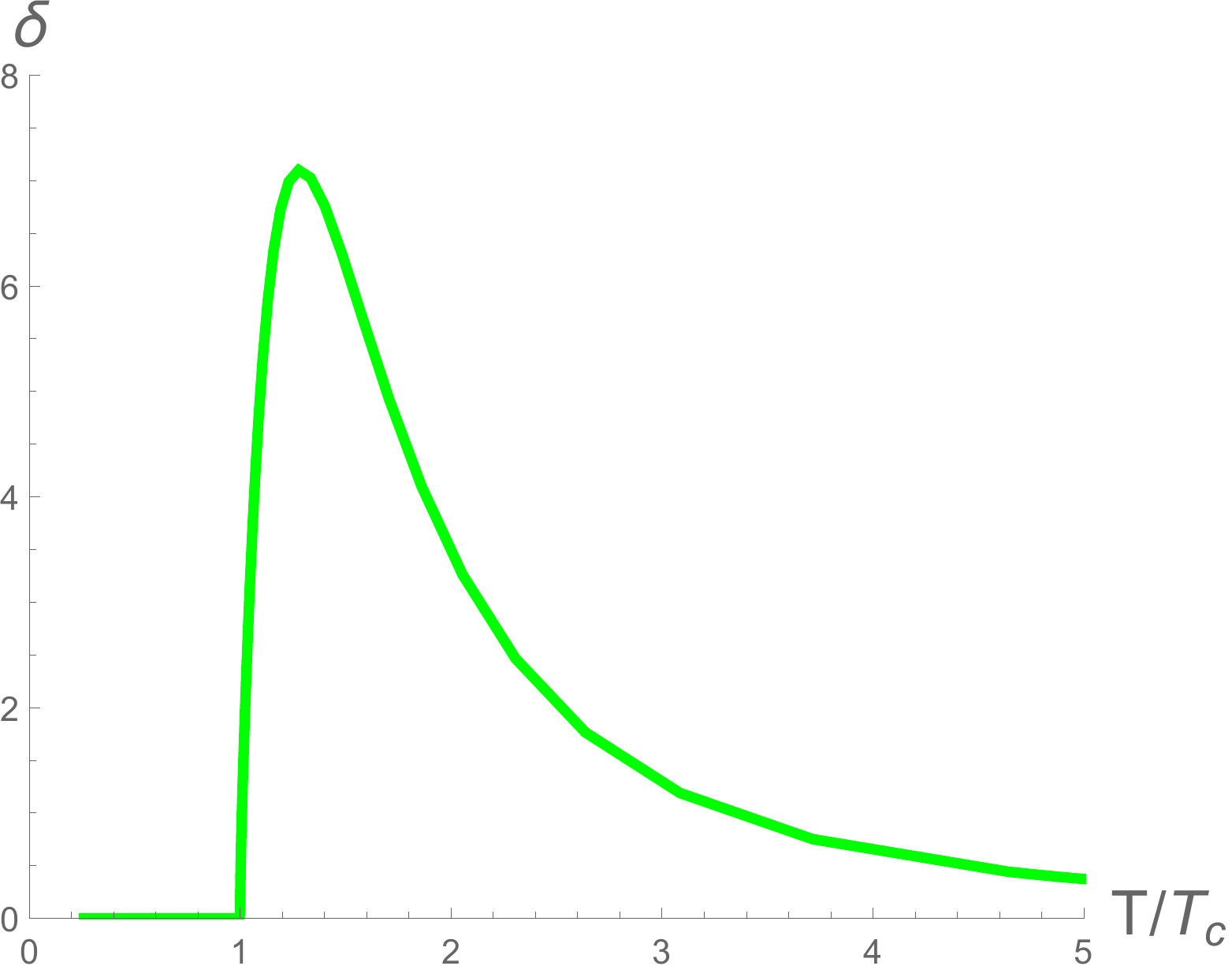}}
\caption{(color online) The temperature dependence of $\delta$, where $\delta \equiv (\frac{2 \varepsilon}{\pi} - \frac{d c}{dt}) /T^4$. It should be noted that $\delta$ is very small but positive in the range $T\leq T_c$.}
\end{figure}

To check the Lloyd's bound Eq.~(\ref{LloydBound}), we need to compare the growth rate of the complexity $d c/dt$ and the energy density $\varepsilon$. Using the thermodynamical relations, we have $\varepsilon = T s - p$, where the pressure $p$ can be calculated as $p = \int_0^T d \bar{T} s(\bar{T})$~\cite{Rougemont:2015wca}. In Fig.~3, we plot the re-scaled difference between $2 \varepsilon/ \pi$ and $d c/dt$. From the figure, we can always see that $\frac{d c}{dt}<\frac{2\varepsilon}{\pi}$ for the three non-conformal cases. This means that the Loyld's bound is always satisfied for the cases we considered but only saturated for the conformal case. This implies that breaking the conformal symmetry will decrease the growth rate of the complexity.

\section{Summary and Discussions}

In this paper, we discuss the behavior of complexity close to crossover/phase transition in the holographic QCD model proposed by Gubser et al~\cite{Gubser:2008ny,Gubser:2008yx}. This holographic model is a Einstein-dilaton system with a specific dilaton potential to fit the lattice results and reproduce the crossover behavior of thermodynamical quantities. Moreover, this simple model can also realize first and second order phase transitions by choosing other values of the parameters within.

Our results show that, for the three non-conformal cases, the re-scaled growth rate of the holographic complexity density exhibits behavior characterizing the corresponding phase structures close to $T_c$: For the crossover case $V_{\rm QCD}$, $\frac{dc/dt}{T^4}$ and its derivative with respect to the temperature are both continuous at $T_c$; For the $1^{\rm st}$ order case $V_{\rm 1st}$, $\frac{dc/dt}{T^4}$ is discontinuous at $T_c$; For the $2^{\rm nd}$ order case $V_{\rm 2nd}$, $\frac{dc/dt}{T^4}$ is continuous at $T_c$ but not its derivative with respect to the temperature; While far beyond the critical point, it approaches constant suggesting that even for the non-conformal cases, $\frac{dc/dt}{T^4}$ may be seen as an effective count of the number of underlying degrees of freedom. And as the re-scaled thermal entropy density $s/T^3$, the growth rate of the holographic complexity density may also be used as a good order parameter to characterize the type of phase transition. So the fast drop or jump of $\frac{dc/dt}{T^4}$ near the critical temperature can be physically understood as a signal of confinement.

Moreover, for the cases we considered, the Lloyd's bound Eq.~(\ref{LloydBound}) is always satisfied but only saturated for the conformal case, which implies that breaking the conformal symmetry will decrease the growth rate of the complexity.

In this work, we only consider the CA conjecture. For the CV conjecture, it is a natural expectation that $\frac{d c}{dt} \sim T s$~\cite{Stanford:2014jda,Brown:2015lvg}. As we have shown in Figs.~2 and 3, the entropy density $s/T^3$ shows a similar behavior as $\frac{d c/dt}{T^4}$. So we can expect that $\frac{d c/dt}{T^4}$ exhibit a similar behavior in the CV conjecture.

We only consider the holographic QCD model proposed by Gubser et al~\cite{Gubser:2008ny,Gubser:2008yx}. Whether our claim is still hold for other holographic models, for example other holographic QCD models Refs.~\cite{Gursoy:2007cb,Gursoy:2007er,Gursoy:2008za,Kiritsis:2009hu,Gursoy:2009jd,
Buchel:2007vy,Buchel:2015saa,Li:2011hp,Cai:2012xh} and holographic superconductor models Refs.~\cite{Hartnoll:2009sz,Herzog:2009xv,McGreevy:2009xe,Horowitz:2010gk,Cai:2015cya}, needs further investigations.

\section*{Acknowledgement}

This work was supported by National Natural Science Foundation of China (No.11605155).


\begin{thebibliography}{99}

\bibitem{Maldacena:1997re}
  J.~M.~Maldacena,
  {\em The Large N limit of superconformal field theories and supergravity},
  Int.\ J.\ Theor.\ Phys.\  {\bf 38}, 1113 (1999)
  [Adv.\ Theor.\ Math.\ Phys.\  {\bf 2}, 231 (1998)]
  [hep-th/9711200].

\bibitem{Gubser:1998bc}
  S.~S.~Gubser, I.~R.~Klebanov and A.~M.~Polyakov,
  {\em Gauge theory correlators from noncritical string theory},
  Phys.\ Lett.\ B {\bf 428}, 105 (1998)
  [hep-th/9802109].

\bibitem{Witten:1998qj}
  E.~Witten,
 {\em Anti-de Sitter space and holography},
  Adv.\ Theor.\ Math.\ Phys.\  {\bf 2}, 253 (1998)
  [hep-th/9802150].

\bibitem{Mateos:2007ay}
  D.~Mateos,
  {\em String Theory and Quantum Chromodynamics},
  Class.\ Quant.\ Grav.\  {\bf 24}, S713 (2007)
  [arXiv:0709.1523 [hep-th]].

\bibitem{Gubser:2009md}
  S.~S.~Gubser and A.~Karch,
  {\em From gauge-string duality to strong interactions: A Pedestrian's Guide},
  Ann.\ Rev.\ Nucl.\ Part.\ Sci.\  {\bf 59}, 145 (2009)
  [arXiv:0901.0935 [hep-th]].

\bibitem{CasalderreySolana:2011us}
  J.~Casalderrey-Solana, H.~Liu, D.~Mateos, K.~Rajagopal and U.~A.~Wiedemann,
  {\em Gauge/String Duality, Hot QCD and Heavy Ion Collisions},
  arXiv:1101.0618 [hep-th].


\bibitem{Hartnoll:2009sz}
  S.~A.~Hartnoll,
  {\em Lectures on holographic methods for condensed matter physics},
  Class.\ Quant.\ Grav.\  {\bf 26}, 224002 (2009)
  [arXiv:0903.3246 [hep-th]].

\bibitem{Herzog:2009xv}
  C.~P.~Herzog,
  {\em Lectures on Holographic Superfluidity and Superconductivity},
  J.\ Phys.\ A {\bf 42}, 343001 (2009)
  [arXiv:0904.1975 [hep-th]].

\bibitem{McGreevy:2009xe}
  J.~McGreevy,
  {\em Holographic duality with a view toward many-body physics},
  Adv.\ High Energy Phys.\  {\bf 2010}, 723105 (2010)
 [arXiv:0909.0518 [hep-th]].

\bibitem{Horowitz:2010gk}
  G.~T.~Horowitz,
 {\em Introduction to Holographic Superconductors},
  Lect.\ Notes Phys.\  {\bf 828}, 313 (2011)
  [arXiv:1002.1722 [hep-th]].

\bibitem{Cai:2015cya}
  R.~G.~Cai, L.~Li, L.~F.~Li and R.~Q.~Yang,
  {\em Introduction to Holographic Superconductor Models},
  Sci.\ China Phys.\ Mech.\ Astron.\  {\bf 58}, no. 6, 060401 (2015)
  [arXiv:1502.00437 [hep-th]].

\bibitem{Banks:2004eb}
  T.~Banks and W.~Fischler,
  {\em The holographic approach to cosmology},
  arXiv:hep-th/0412097.

\bibitem{Swingle:2009bg}
  B.~Swingle,
 {\em Entanglement Renormalization and Holography},
  Phys.\ Rev.\ D {\bf 86}, 065007 (2012)
  [arXiv:0905.1317 [cond-mat.str-el]].

\bibitem{Swingle:2012wq}
  B.~Swingle,
  {\em Constructing holographic spacetimes using entanglement renormalization},
  arXiv:1209.3304 [hep-th].

\bibitem{Qi:2013caa}
  X.~L.~Qi,
  {\em Exact holographic mapping and emergent space-time geometry},
  arXiv:1309.6282 [hep-th].


\bibitem{Susskind:2014rva}
  L.~Susskind,
  {\em Computational Complexity and Black Hole Horizons},
  [Fortsch.\ Phys.\  {\bf 64}, 24 (2016)]
  Addendum: Fortsch.\ Phys.\  {\bf 64}, 44 (2016)
  [arXiv:1403.5695 [hep-th], arXiv:1402.5674 [hep-th]].

\bibitem{Stanford:2014jda}
  D.~Stanford and L.~Susskind,
  {\em Complexity and Shock Wave Geometries},
  Phys.\ Rev.\ D {\bf 90}, no. 12, 126007 (2014)
  [arXiv:1406.2678 [hep-th]].


\bibitem{Brown:2015bva}
  A.~R.~Brown, D.~A.~Roberts, L.~Susskind, B.~Swingle and Y.~Zhao,
  {\em Holographic Complexity Equals Bulk Action?},
  Phys.\ Rev.\ Lett.\  {\bf 116}, no. 19, 191301 (2016)
  [arXiv:1509.07876 [hep-th]].

\bibitem{Brown:2015lvg}
  A.~R.~Brown, D.~A.~Roberts, L.~Susskind, B.~Swingle and Y.~Zhao,
  {\em Complexity, action, and black holes},
  Phys.\ Rev.\ D {\bf 93}, no. 8, 086006 (2016)
  [arXiv:1512.04993 [hep-th]].

\bibitem{Lloyd:2000}
  S. Lloyd,
  {\em Ultimate physical limits to computation},
  Nature {\bf 406}, 1047 (2000).

\bibitem{Alishahiha:2015rta}
  M.~Alishahiha,
  {\em Holographic Complexity},
  Phys.\ Rev.\ D {\bf 92}, no. 12, 126009 (2015)
  [arXiv:1509.06614 [hep-th]].

\bibitem{Momeni:2016ekm}
  D.~Momeni, S.~A.~H.~Mansoori and R.~Myrzakulov,
  {\em Holographic Complexity in Gauge/String Superconductors},
  Phys.\ Lett.\ B {\bf 756}, 354 (2016)
  [arXiv:1601.03011 [hep-th]].

\bibitem{Cai:2016xho}
  R.~G.~Cai, S.~M.~Ruan, S.~J.~Wang, R.~Q.~Yang and R.~H.~Peng,
  {\em Action growth for AdS black holes},
  JHEP {\bf 1609}, 161 (2016)
  [arXiv:1606.08307 [gr-qc]].


\bibitem{Brown:2016wib}
  A.~R.~Brown, L.~Susskind and Y.~Zhao,
  {\em Quantum Complexity and Negative Curvature},
  Phys.\ Rev.\ D {\bf 95}, no. 4, 045010 (2017)
  [arXiv:1608.02612 [hep-th]].


\bibitem{Couch:2016exn}
  J.~Couch, W.~Fischler and P.~H.~Nguyen,
  {\em Noether charge, black hole volume, and complexity},
  JHEP {\bf 1703}, 119 (2017)
  [arXiv:1610.02038 [hep-th]].

\bibitem{Yang:2016awy}
  R.~Q.~Yang,
  {\em Strong energy condition and complexity growth bound in holography},
  Phys.\ Rev.\ D {\bf 95}, no. 8, 086017 (2017)
  [arXiv:1610.05090 [gr-qc]].


\bibitem{Chapman:2016hwi}
  S.~Chapman, H.~Marrochio and R.~C.~Myers,
  {\em Complexity of Formation in Holography},
  JHEP {\bf 1701}, 062 (2017)
  [arXiv:1610.08063 [hep-th]].

\bibitem{Carmi:2016wjl}
  D.~Carmi, R.~C.~Myers and P.~Rath,
  {\em Comments on Holographic Complexity},
  JHEP {\bf 1703}, 118 (2017)
  [arXiv:1612.00433 [hep-th]].


\bibitem{Pan:2016ecg}
  W.~J.~Pan and Y.~C.~Huang,
  {\em Holographic complexity and action growth in massive gravities},
  Phys.\ Rev.\ D {\bf 95}, no. 12, 126013 (2017)
  [arXiv:1612.03627 [hep-th]].

\bibitem{Brown:2017jil}
  A.~R.~Brown and L.~Susskind,
  {\em The Second Law of Quantum Complexity},
  arXiv:1701.01107 [hep-th].


\bibitem{Kim:2017lrw}
  R.~Q.~Yang, C.~Niu and K.~Y.~Kim,
  {\em Surface Counterterms and Regularized Holographic Complexity},
  JHEP {\bf 1709}, 042 (2017)
  [arXiv:1701.03706 [hep-th]].

\bibitem{Cai:2017sjv}
  R.~G.~Cai, M.~Sasaki and S.~J.~Wang,
  {\em Action growth of charged black holes with a single horizon},
  Phys.\ Rev.\ D {\bf 95}, no. 12, 124002 (2017)
  [arXiv:1702.06766 [gr-qc]].


\bibitem{Alishahiha:2017hwg}
  M.~Alishahiha, A.~Faraji Astaneh, A.~Naseh and M.~H.~Vahidinia,
  {\em On complexity for $F(R)$ and critical gravity},
  JHEP {\bf 1705}, 009 (2017)
  [arXiv:1702.06796 [hep-th]].

\bibitem{Bakhshaei:2017qud}
  E.~Bakhshaei, A.~Mollabashi and A.~Shirzad,
  {\em Holographic Subregion Complexity for Singular Surfaces},
  Eur.\ Phys.\ J.\ C {\bf 77}, no. 10, 665 (2017)
  [arXiv:1703.03469 [hep-th]].

\bibitem{Tao:2017fsy}
  J.~Tao, P.~Wang and H.~Yang,
  {\em Testing holographic conjectures of complexity with Born-Infeld black holes},
  Eur.\ Phys.\ J.\ C {\bf 77}, no. 12, 817 (2017)
  [arXiv:1703.06297 [hep-th]].


\bibitem{Guo:2017rul}
  W.~D.~Guo, S.~W.~Wei, Y.~Y.~Li and Y.~X.~Liu,
  {\em Complexity growth rates for AdS black holes in massive gravity and $f(R)$ gravity},
  arXiv:1703.10468 [gr-qc].


\bibitem{Zangeneh:2017tub}
  M.~Kord Zangeneh, Y.~C.~Ong and B.~Wang,
  {\em Entanglement Entropy and Complexity for One-Dimensional Holographic Superconductors},
  Phys.\ Lett.\ B {\bf 771}, 235 (2017)
  [arXiv:1704.00557 [hep-th]].


\bibitem{Alishahiha:2017cuk}
  M.~Alishahiha and A.~Faraji Astaneh,
  {\em Holographic Fidelity Susceptibility},
  Phys.\ Rev.\ D {\bf 96}, no. 8, 086004 (2017)
  [arXiv:1705.01834 [hep-th]].


\bibitem{Abad:2017cgl}
  F.~J.~G.~Abad, M.~Kulaxizi and A.~Parnachev,
  {\em On Complexity of Holographic Flavors},
  arXiv:1705.08424 [hep-th].


\bibitem{Reynolds:2017lwq}
  A.~Reynolds and S.~F.~Ross,
  {\em Complexity in de Sitter Space},
  Class.\ Quant.\ Grav.\  {\bf 34}, no. 17, 175013 (2017)
  [arXiv:1706.03788 [hep-th]].


\bibitem{Hashimoto:2017fga}
  K.~Hashimoto, N.~Iizuka and S.~Sugishita,
  {\em Time evolution of complexity in Abelian gauge theories},
  Phys.\ Rev.\ D {\bf 96}, no. 12, 126001 (2017)
  [arXiv:1707.03840 [hep-th]].


\bibitem{Nagasaki:2017kqe}
  K.~Nagasaki,
  {\em Complexity of $AdS_5$ black holes with a rotating string},
  arXiv:1707.08376 [hep-th].

\bibitem{Miao:2017quj}
  Y.~G.~Miao and L.~Zhao,
  {\em Complexity/Action duality of the shock wave geometry in a massive gravity theory},
  arXiv:1708.01779 [hep-th].

\bibitem{Ge:2017rak}
  X.~H.~Ge and B.~Wang,
  {\em Quantum computational complexity, Einstein's equations and accelerated expansion of the Universe},
  arXiv:1708.06811 [hep-th].

\bibitem{Ghodrati:2017roz}
  M.~Ghodrati,
  {\em Complexity growth in massive gravity theories, the effects of chirality, and more},
  Phys.\ Rev.\ D {\bf 96}, no. 10, 106020 (2017)
  [arXiv:1708.07981 [hep-th]].


\bibitem{Qaemmaqami:2017lzs}
  M.~M.~Qaemmaqami,
  {\em On Complexity Growth in Minimal Massive 3D Gravity},
  arXiv:1709.05894 [hep-th].


\bibitem{Carmi:2017jqz}
  D.~Carmi, S.~Chapman, H.~Marrochio, R.~C.~Myers and S.~Sugishita,
  {\em On the Time Dependence of Holographic Complexity},
  JHEP {\bf 1711}, 188 (2017)
  [arXiv:1709.10184 [hep-th]].

\bibitem{Kim:2017qrq}
  K.~Y.~Kim, C.~Niu, R.~Q.~Yang and C.~Y.~Zhang,
  {\em Comparison of holographic and field theoretic complexities by time dependent thermofield double states},
  arXiv:1710.00600 [hep-th].

\bibitem{Cottrell:2017ayj}
  W.~Cottrell and M.~Montero,
  {\em Complexity is Simple},
  arXiv:1710.01175 [hep-th].


\bibitem{Sebastiani:2017rxr}
  L.~Sebastiani, L.~Vanzo and S.~Zerbini,
  {\em Action growth for black holes in modified gravity},
  arXiv:1710.05686 [hep-th].

\bibitem{Moosa:2017yvt}
  M.~Moosa,
  {\em Evolution of Complexity Following a Global Quench},
  arXiv:1711.02668 [hep-th].

\bibitem{Reynolds:2017jfs}
  A.~P.~Reynolds and S.~F.~Ross,
  {\em Complexity of the AdS Soliton},
  arXiv:1712.03732 [hep-th].


\bibitem{HosseiniMansoori:2017tsm}
  S.~A.~Hosseini Mansoori and M.~M.~Qaemmaqami,
  {\em Complexity Growth, Butterfly Velocity and Black hole Thermodynamics},
  arXiv:1711.09749 [hep-th].

\bibitem{Gubser:2008ny}
  S.~S.~Gubser and A.~Nellore,
  {\em Mimicking the QCD equation of state with a dual black hole},
  Phys.\ Rev.\ D {\bf 78}, 086007 (2008)
  [arXiv:0804.0434 [hep-th]].

\bibitem{Gubser:2008yx}
  S.~S.~Gubser, A.~Nellore, S.~S.~Pufu and F.~D.~Rocha,
  {\em Thermodynamics and bulk viscosity of approximate black hole duals to finite temperature quantum chromodynamics},
  Phys.\ Rev.\ Lett.\  {\bf 101}, 131601 (2008)
  [arXiv:0804.1950 [hep-th]].


\bibitem{Borsanyi:2012cr}
  S.~Borsanyi, G.~Endrodi, Z.~Fodor, S.~D.~Katz, S.~Krieg, C.~Ratti and K.~K.~Szabo,
  {\em QCD equation of state at nonzero chemical potential: continuum results with physical quark masses at order $mu^2$},
  JHEP {\bf 1208}, 053 (2012)
 [arXiv:1204.6710 [hep-lat]].

\bibitem{Finazzo:2014cna}
  S.~I.~Finazzo, R.~Rougemont, H.~Marrochio and J.~Noronha,
  {\em Hydrodynamic transport coefficients for the non-conformal quark-gluon plasma from holography},
  JHEP {\bf 1502}, 051 (2015)
  [arXiv:1412.2968 [hep-ph]].

\bibitem{Janik:2015waa}
  R.~A.~Janik, G.~Plewa, H.~Soltanpanahi and M.~Spalinski,
  {\em Linearized nonequilibrium dynamics in nonconformal plasma},
  Phys.\ Rev.\ D {\bf 91}, no. 12, 126013 (2015)
 [arXiv:1503.07149 [hep-th]].

\bibitem{Janik:2016btb}
  R.~A.~Janik, J.~Jankowski and H.~Soltanpanahi,
  {\em Quasinormal modes and the phase structure of strongly coupled matter},
  JHEP {\bf 1606}, 047 (2016)
  [arXiv:1603.05950 [hep-th]].

\bibitem{DeWolfe:2010he}
  O.~DeWolfe, S.~S.~Gubser and C.~Rosen,
  {\em A holographic critical point},
  Phys.\ Rev.\ D {\bf 83}, 086005 (2011)
 [arXiv:1012.1864 [hep-th]].

\bibitem{Zhang:2016rcm}
  S.~J.~Zhang,
  {\em Holographic entanglement entropy close to crossover/phase transition in strongly coupled systems},
  Nucl.\ Phys.\ B {\bf 916}, 304 (2017)
  [arXiv:1608.03072 [hep-th]].

\bibitem{Knaute:2017lll}
  J.~Knaute and B.~K$\ddot{a}$mpfer,
  {\em Holographic Entanglement Entropy in the QCD Phase Diagram with a Critical Point},
  Phys.\ Rev.\ D {\bf 96}, no. 10, 106003 (2017)
  [arXiv:1706.02647 [hep-ph]].


\bibitem{Breitenlohner:1982bm}
  P.~Breitenlohner and D.~Z.~Freedman,
  {\em Positive Energy in anti-De Sitter Backgrounds and Gauged Extended Supergravity},
  Phys.\ Lett.\ B {\bf 115}, 197 (1982).

\bibitem{Breitenlohner:1982jf}
  P.~Breitenlohner and D.~Z.~Freedman,
  {\em Stability in Gauged Extended Supergravity},
  Annals Phys.\  {\bf 144}, 249 (1982).


\bibitem{Lehner:2016vdi}
  L.~Lehner, R.~C.~Myers, E.~Poisson and R.~D.~Sorkin,
  {\em Gravitational action with null boundaries},
  Phys.\ Rev.\ D {\bf 94}, no. 8, 084046 (2016)
  [arXiv:1609.00207 [hep-th]].


\bibitem{Rougemont:2015wca}
  R.~Rougemont, A.~Ficnar, S.~Finazzo and J.~Noronha,
  {\em Energy loss, equilibration, and thermodynamics of a baryon rich strongly coupled quark-gluon plasma},
  JHEP {\bf 1604}, 102 (2016)
  [arXiv:1507.06556 [hep-th]].


\bibitem{Gursoy:2007cb}
  U.~Gursoy and E.~Kiritsis,
  {\em Exploring improved holographic theories for QCD: Part I},
  JHEP {\bf 0802}, 032 (2008)
  [arXiv:0707.1324 [hep-th]].

\bibitem{Gursoy:2007er}
  U.~Gursoy, E.~Kiritsis and F.~Nitti,
  {\em Exploring improved holographic theories for QCD: Part II},
  JHEP {\bf 0802}, 019 (2008)
  [arXiv:0707.1349 [hep-th]].

\bibitem{Gursoy:2008za}
  U.~Gursoy, E.~Kiritsis, L.~Mazzanti and F.~Nitti,
  {\em Holography and Thermodynamics of 5D Dilaton-gravity},
  JHEP {\bf 0905}, 033 (2009)
 [arXiv:0812.0792 [hep-th]].

\bibitem{Kiritsis:2009hu}
  E.~Kiritsis,
  {\em Dissecting the string theory dual of QCD},
  Fortsch.\ Phys.\  {\bf 57}, 396 (2009)
  [arXiv:0901.1772 [hep-th]].

\bibitem{Gursoy:2009jd}
  U.~Gursoy, E.~Kiritsis, L.~Mazzanti and F.~Nitti,
  {\em Improved Holographic Yang-Mills at Finite Temperature: Comparison with Data},
  Nucl.\ Phys.\ B {\bf 820}, 148 (2009)
 [arXiv:0903.2859 [hep-th]].

\bibitem{Buchel:2007vy}
  A.~Buchel, S.~Deakin, P.~Kerner and J.~T.~Liu,
  {\em Thermodynamics of the N=2* strongly coupled plasma},
  Nucl.\ Phys.\ B {\bf 784}, 72 (2007)
 [hep-th/0701142].

\bibitem{Buchel:2015saa}
  A.~Buchel, M.~P.~Heller and R.~C.~Myers,
  {\em Equilibration rates in a strongly coupled nonconformal quark-gluon plasma},
  Phys.\ Rev.\ Lett.\  {\bf 114}, no. 25, 251601 (2015)
  [arXiv:1503.07114 [hep-th]].

\bibitem{Li:2011hp}
  D.~Li, S.~He, M.~Huang and Q.~S.~Yan,
  {\em Thermodynamics of deformed AdS$_5$ model with a positive/negative quadratic correction in graviton-dilaton system},
  JHEP {\bf 1109}, 041 (2011)
  [arXiv:1103.5389 [hep-th]].

\bibitem{Cai:2012xh}
  R.~G.~Cai, S.~He and D.~Li,
  {\em A hQCD model and its phase diagram in Einstein-Maxwell-Dilaton system},
  JHEP {\bf 1203}, 033 (2012)
  [arXiv:1201.0820 [hep-th]].


%
%
%
%
%
%
%
%
%
%
%
%
%
%
%
%
%
%
%
%
%
%
%
%
%
%
%
%
%
%
%
\end{thebibliography}
\end{document}